\documentclass{article}
\usepackage{graphicx}  % standard LaTeX graphics tool;
\usepackage{amsmath}   % For getting proper math eqns
\usepackage[compress]{cite}
\usepackage{amssymb}   % Used for getting various symbols eg. gtrsim, lesssim etc.
\usepackage{bm} % For bold math (esp of lower greek letters)
\usepackage{dcolumn}% Align table columns on decimal point
\usepackage{color}
\usepackage{mathrsfs}
\usepackage{amsfonts}
\usepackage{varioref}
\usepackage{textcomp}

\RequirePackage[colorlinks,citecolor=blue,urlcolor=magenta,linkcolor=blue]{hyperref}
\allowdisplaybreaks

\labelformat{section}{Section #1} 
\labelformat{subsection}{Section #1} 
\labelformat{subsubsection}{Section #1}
\labelformat{subsubsubsection}{Section #1}
\labelformat{equation}{Eq.~(#1)} 
\labelformat{figure}{Fig.~#1} 
\labelformat{subfigure}{Fig.~\thefigure#1} 
\labelformat{table}{Table~#1} 
\labelformat{appendix}{Appendix #1}
\newcounter{mnotecount}[section]

\renewcommand{\themnotecount}{\thesection.\arabic{mnotecount}}
\newcommand{\cred}{{\color{red}}}
\newcommand{\mnotex}[1]%{}
{\protect{\stepcounter{mnotecount}}$^{\mbox{\footnotesize
$%\!\!\!\!\!\!\,
\bullet$\themnotecount}}$ \marginpar{
\raggedright\small\em
$\!\!\!\!\!\!\,\bullet$\themnotecount: #1} }
\begin{document}
\title{On electrogravity duality and black hole with global monopole }
%\author{Xi\'an O. Camanho }
%\affiliation{Max-Planck-Institut f\"ur Gravitationsphysik, Albert-Einstein-Institut, 1447thebibliography6 Golm, Germany}
\author{Chiranjeeb Singha\footnote{\href{mailto:chiranjeeb.singha@iucaa.in}{chiranjeeb.singha@iucaa.in}}$~^{1}$, and Naresh Dadhich\footnote{\href{mailto:nkd@iucaa.in}{nkd@iucaa.in}}$~^{1,\;2}$\\
$^{1}$\small{Inter-University Centre for Astronomy \& Astrophysics, Post Bag 4, Pune 411 007, India}\\ 
$^{2}$\small{Astrophysics Research Centre, School of Mathematics,
Statistics and Computer Science,}\\
\small{ University of KwaZulu-Natal,
Private Bag X54001, Durban 4000, South Africa}
}
%\author{Alfred Molina}
%\affiliation{Departament de F\'{\i}sica Fonamental, Universitat de Barcelona, Spain}

\date{\today}
%%%%%%%%%%%%%%%%%%%%%%%%%%%%%%%%%%%%%%%%%%%%%%%%%%%%%%%%%%%%%%%%%%%%%%%%%%%%%%%%%%%%%%%%%%%%%%%%%%%
%%%%%%%%%%%%%%%%%%%%%%%%%%%%%%%%%%%%%%%%%%%%%%%%%%%%%%%%%%%%%%%%%%%%%%%%%%%%%%%%%%%%%%%%%%%%%%%%%%%
\maketitle
%%%%%%%%%%%%%%%%%%%%%%%%%%%%%%%%%%%%%%%%%%%%%%%%%%%%%%%%%%%%%%%%%%%%%%%%%%%%%%%%%%%%%%%%%%%%%%%%%%%
%%%%%%%%%%%%%%%%%%%%%%%%%%%%%%%%%%%%%%%%%%%%%%%%%%%%%%%%%%%%%%%%%%%%%%%%%%%%%%%%%%%%%%%%%%%%%%%%%%%

%%%%%%%%%%%%%%%%%%%%%%%%%%%%%%%%%%%%%%%%%%%%%%%%%%%%%%%%%%%%%%%%%%%%%%%%%%%%%%%%%%%%%%%%%%%%%%%%%%%
%%%%%%%%%%%%%%%%%%%%%%%%%%%%%%%%%%%%%%%%%%%%%%%%%%%%%%%%%%%%%%%%%%%%%%%%%%%%%%
%%%%%%%%%%%%%%%%%%%%%
\begin{abstract}

By resolving the Riemann curvature into electric and magnetic parts, Einstein's equation can accordingly be written in terms of electric (active and passive) and magnetic parts. The electrogravity duality is defined by the interchange of active and passive parts. It turns out that in static and stationary spacetimes, there is a subset of the equations (that identifies the effective vacuum equation) that is sufficient to yield the vacuum solution. In spherically symmetric spacetime, the electrograv dual of the effective equation solves to give the Schwarzschild black hole with a global monopole. Interestingly, this is not so for axial symmetry,  where the Kerr vacuum solution turns out to be electrograv self-dual. However, in the asymptotic limit where the effect of rotation dies out, the situation reverts to the static case, admitting a global monopole. This is also what follows when we apply the Newman-Janis transformation to the static black hole with a global monopole.   

\end{abstract}
%%%%%%%%%%%%%%%%%%%%%%%%%%%%%%%%%%%%%%%%%%%%%%%%%%%%%%%%%%%%%%%%%%%%%%%%%%%%%%%%%%%%%%%%%%%%%%%%%%%%%%%%%%%%%%%%%%%%%%%%%%%%%%%%%%%%%%%%%%%%%%%%%%%%%%%%%%%%%%%%%%%%%%%%%%%%%%%%%%%%%%%%%%%%%%%%%%%%%%%%%%%%%%%%%%%%%%%%%%%%%%%%%%%%%

\section{Introduction}

A fundamental aspect of understanding the Einstein field equations lies in analyzing the curvature of spacetime as encoded in the Riemann tensor. A particularly insightful approach is to decompose the Riemann tensor into its electric and magnetic parts with respect to a timelike unit vector field \( u^a \), representing the 4-velocity of an observer \cite{Dadhich:1999eh, Dadhich:2000df, Bose:1999ba, Bose:1999ke, Dadhich:2000jk, Dadhich:1999eg, Dadhich:1999gj, Nouri-Zonoz:1998whb}. This decomposition, analogous to the electromagnetic field tensor split, yields a clearer physical interpretation of spacetime geometry.

We define the \emph{active} electric part by $E_{ac}= R_{abcd} u^b u^d$,  \emph{passive} electric part by $\tilde{E}_{ac}= {}^*R^*_{abcd} u^b u^d$, and the magnetic part by $H_{ac} = {}^*R_{abcd} u^b u^d$ where a star denotes the Hodge dual. The electromagnetic parts are by definition the spatial second rank tensors orthogonal to $u^a$. The former two are symmetric, accounting for 12 components, while the latter is trace-free, making up the remaining 8 of the Riemann components. These will play a central role in characterizing the geometry of spacetime, especially in vacuum and electrovacuum configurations.

In general, the vacuum Einstein equation exhibits symmetry between the active and passive electric parts. However, a degeneracy emerges in the field equations in highly symmetric spacetimes such as those with spherical or axial symmetry. For instance, in a static, spherically symmetric spacetime, if the Einstein tensor components satisfy $G^t_{\ t} = G^r_{\ r} = 0$, that implies $G^\theta_{\ \theta} = 0$, giving the vacuum solution. That is, a subset of equations suffices to determine the full solution. This redundancy is what we probe through the electrogravity duality transformation. 

The electrograv duality is characterized by the interchange of active and passive electric parts, ${E}_{ab} \leftrightarrow \tilde{E}_{ab}$, which is equivalent to the interchange of Ricci and Einstein tensors. This is because the former is a trace of Riemann while the latter is that of its double dual. Similarly, under electrograv duality, scalar curvature $R$ goes to $-R$.   

The solution of the electrograv dual equation for spherically symmetric spacetime introduces a global monopole on the Schwarzschild black hole. The source of global monopole turns out to be the topological defect, solid angle deficit \cite{Dadhich:1997ue}. This, however, does not happen for axial symmetry, where the Kerr vacuum solution is electrograv self-dual; i.e., the dual equation also admits the same Kerr solution. This is perhaps because, unlike spherical symmetry, it is not possible to effect a solid angle deficit in axial symmetry, except in the asymptotic limit of large $r$. In that case, the effect of rotation dies out, and the spacetime effectively tends to have spherical symmetry that could very well harbor a solid angle deficit and consequently a global monopole. This is precisely what is borne out by the Newman-Janis transformation of the Schwarzschild black hole with a global monopole. This would, however, not in general be a solution to the dual vacuum equation except in the asymptotic limit.

The topological defects are characterized by the vanishing of focusing (or active gravitational mass, $\rho + 3p$) density. It turns out \cite{Dadhich:1997ue} that this could be achieved by a geometric ansatz by writing $n$-dimensional Euclidean or Minkowski distance proportional to the corresponding $(n-1)$-dimensional distance. This will generate a topological defect like a solid angle deficit in $(n-1)$-dimensional space or spacetime, and consequently a global monopole or global texture. 

To systematically explore the electrograv duality, we introduce three physically insightful energy density constructs: the energy density as measured by a timelike observer: $\rho = T_{ab} u^a u^b $, the null energy density, $ \rho_n = R_{ab} n^a n^b $ relative a null vector, and the focusing energy density, $ \rho_f = (T_{ab} - 1/2 ~Tg_{ab})u^a u^b $ \cite{Dadhich:2001fu}.

The effective vacuum equation thus reads as $\rho=0, \; \rho_n = 0$ while its dual will be $\rho_f = 0,\; \rho_n =0$. Note that $\rho$ and  $\rho_f$ are dual to each other while $\rho_n$ is self-dual. Clearly, the effective vacuum equation is not invariant under the electrogravity duality transformation, indicated by the interchange, $ E_{ab} \leftrightarrow \tilde{E}_{ab} $. This asymmetry gives rise to two distinct solutions. The standard vacuum condition, $ \rho = \rho_n =0$ corresponds to the familiar black hole spacetimes such as Schwarzschild and Kerr. In contrast, the electrograv dual condition $ \rho_f = \rho_n =0$ leads to a different solution in static spacetime with nontrivial topological structure, such as black holes with solid angle deficit representing asymptotically a global monopole  \cite{PhysRevLett.63.341, TeixeiraFilho:2001fc, Mazur:1990ak,  Achucarro:2000td, Harari:1990cz}. That is the solution of the dual vacuum equation, the Schwarzschild black hole with a global monopole charge. On the other hand, the Kerr solution is electrograv self-dual. However, as alluded to above, Kerr black hole harbor a global monopole asymptotically.   

For studying the topological defects like global monopole, the electrograv duality offers a new impetus and insight. That is, it provides an insightful framework for introducing topological defects in vacuum and electrovac spacetimes, thereby enriching the solution space of Einstein's equations and offering new avenues for physical exploration.  

We shall also show that the dual spacetime metric indeed represents asymptotically the stress tensor of a global monopole produced by the spontaneous breaking of $O(3)$ symmetry into $U(1)$. 

The paper is organized as follows: In Section \ref{Electrogrvity}, we decompose the Riemann curvature tensor relative to a timelike unit vector and write the vacuum equation and its dual (henceforth by dual we will mean electrograv dual) in terms of electric and magnetic parts for spherical and axially symmetric spacetimes and then obtain the corresponding solutions. By performing the Newman-Janis algorithm on the Schwarzschild black hole with a global monopole \ref{global}, and by following \cite{PhysRevLett.63.341}, we shall demonstrate that the dual spacetime does indeed asymptotically describe a black hole with a global monopole. Finally, we conclude in \ref{discussion} with a discussion.

\paragraph{\textit{Notations and Conventions}} 
Throughout this paper, we adopt the positive signature convention. Specifically, in $(3+1)$-dimensional Cartesian coordinates, the Minkowski metric is given by $\text{diag}(-1, +1, +1,\\ +1)$. We work in the natural units, setting $G = c = 1$. The notation $f'(x)$ denotes the first derivative of the function $f(x)$ with respect to the variable $x$.

%%%%%%%%%%%%%%%%%%%%%%%%%%%%%%%%%%%%%%%%%%%%%%%%%%%%%%%%%%%%%%%%%%%%%%%%%%%%%%%%%%%%%%%%%%%%%%%%%%%%%%%%%%%%%%%%%%%%%%%%%%%%%%%%%%%%%%%%%%%%%%%%%%%%%%%%%%%%%%%%%%%%%%%%%%%%%%%%%%%%%%%%%%%%%%%%%%%%%%%%%%%%%%%%%%%%%%%%%%%%%%%%%%%%

\section{Electrogravity duality} \label{Electrogrvity} 

We shall begin by resolving the Riemann tensor into electric and magnetic parts, consisting  of six components, each of active and passive electric parts symmetric tensors, $E_{ab}, \tilde E_{ab}$ 
and the trace-free magnetic part $H_{ab}$ accounting for the remaining eight components. That will be followed by the effective vacuum equation and its dual, and their solutions. We shall also illustrate the application of the Newman-Janis algorithm to obtain the Kerr black hole with a global monopole metric from its Schwarzschild counterpart.

\subsection{Electromagnetic decomposition of Riemann and Ricci}
 We begin with the decomposition of the Riemann curvature relative to a timelike unit vector as follows: $E_{ac}= R_{abcd}u^b u^d, \quad
{\tilde E}_{ac}= *R*_{abcd}u^b u^d~, \quad
H_{ac}= *R_{abcd}u^b u^d = H_{(ac)} + H_{[ac]}~,$ 
where
$H_{(ac)}= *C_{abcd}u^b u^d, \quad
H_{[ac]}= \frac{1}{2} \eta _{abce} R^e_d u^b u^d$ (note that $R*_{abcd}u^b u^d = -H_{ac}$).
Here, $C_{abcd}$ represents the Weyl conformal curvature, and $\eta _{abcd}$ denotes the 4-dimensional volume element. The following relations hold: $ E_{ab} = E_{ba}, \quad {\tilde E}_{ab} = {\tilde {E}}_{ba},$ $(E_{ab}, {\tilde {E}}_{ab},H_{ab}) u^b  = 0,$  and
$ H=H^a_a=0.$
%%%%%%%%%%%%%%%%%%%%%%%%%%%%%%%%%%%%%%%%%%%%%%%%%%%%%%
%In Ref. \cite{Dadhich:1999eh}, $E_{ab}$ and ${\tilde E}_{ab}$ are respectively termed as the active and passive electric parts. Using the above decomposition, the Ricci tensor can be written as
%%%%%%%%%%%%%%%%%%%%%%%%%%%%%%%%%%%%%%%%%%%%%%%%%%%%%%

Now we can write the Ricci tensor in terms of electromagnetic parts as follows: 
\begin{equation}
R_a^b = E_a^b +{\tilde {E}}^b_a+ (E + {\tilde {E}})u_a u^b -
{\tilde {E}}g^b_a+\frac{1}{2}
\left( \eta _{amnc} H^{mn}u^b u^c + \eta ^{bmnc}H_{mn}u_a u_c \right),
\end{equation}
%%%%%%%%%%%%%%%%%%%%%%%%%%%%%%%%%%%%%%%%%%%%%%%%%%%%%%%%
where $E=E^a_a$ and ${\tilde {E}}={\tilde {E}}^a_a$. The non-vanishing components of $E_{ab}, \tilde E_{ab}, H_{ab}$ are for spherical and axially symmetric spacetimes are given by  
%%%%%%%%%%%%%%%%%%%%%%%%%%%%%%%%%%%%%%%%%%%%%%%%%%%%%%%%%%%%%%%%%%  
\begin{equation}  
\begin{split}  
    &E_{rr} = R_{rtrt}, \quad E_{\theta \theta} = R_{\theta t \theta t}, \quad E_{\phi \phi} = R_{\phi t \phi t},\\
    &\tilde{E}_{rr} = R_{\theta \phi \theta \phi},   \quad \tilde{E}_{\theta \theta} = R_{r\phi r\phi} , \quad \tilde{E}_{\phi \phi}= R_{r\theta r\theta},\\
      &H_{rr} = R_{t r \theta \phi}, \quad H_{\theta \theta} = R_{t \phi r \theta} = H_{\phi \phi}.
\end{split}  
\end{equation}  
%%%%%%%%%%%%%%%%%%%%%%%%%%%%%%%%%%%%%%%%%%%%%%%%%%%%%%%%%%
Note that the antisymmetric part of $H_{ab}$ is identically zero for the spherical/axial symmetric spacetime.
%%%%%%%%%%%%%%%%%%%%%%%%%%%%%%%%%%%%%%%%%%%%%%%%%%%%%%%%%%%%%%%%%%%%%%%%%%%%%%%%%%%%%%%%%%%%%%%%%%%%%%%%%%%%%%%%%%%%%%%%%%%%%%%%%%%%%%%%%%%%%%%%%%%%%%%%%%%%%%%%%%%%%%%%%%%%%%%%%%%%%%%%%%%%%%%%%%%%%%%%%%%%%%%%%%%%%%%%%%%%%%%%%%%  

\subsection{Vacuum equation and its dual}
The vacuum equation, $R_{ab} = 0$ is in general equivalent to  
%%%%%%%%%%%%%%%%%%%%%%%%%%%%%%%%%%%%%%%%%%%%%%%%%%%%%%%%%%%%
    \begin{equation}
     E ~ or ~ {\tilde E} = 0,~ H_{[ab]} = 0 = E_{ab} + {\tilde E}_{ab}
     \end{equation}
%%%%%%%%%%%%%%%%%%%%%%%%%%%%%%%%%%%%%%%%%%%%%%%%%%%%%%%%%%%%%     
\noindent which is symmetric in $E_{ab}$ and ${\tilde E}_{ab}$.
    
\noindent We define the duality transformation as \cite{Dadhich:1999eh}, 
%%%%%%%%%%%%%%%%%%%%%%%%%%%%%%%%%%%%%%%%%%%%%%%%%%%%%%%%%%%%%
\begin{equation}
      E_{ab} \longleftrightarrow {\tilde E}_{ab}.
      \end{equation}
%%%%%%%%%%%%%%%%%%%%%%%%%%%%%%%%%%%%%%%%%%%%%%%%%%%%%%%%%%     
\noindent The above duality transformation preserves the structure of the vacuum Einstein equations and represents a symmetry of the Einstein-Hilbert action, since the scalar curvature $R$ remains invariant \cred{modulo the sign} under this transformation. Specifically, the duality induces the interchange $
R_{ab} \longleftrightarrow G_{ab},
$
where $ G_{ab} $ is the Einstein tensor. This is because the Ricci tensor is the contraction of the Riemann, while the Einstein is that of its double dual.

\noindent Let's define the relevant energy densities as follows:  
%%%%%%%%%%%%%%%%%%%%%%%%%%%%%%%%%%%%%%%%%%%%%%%%%%%%%%%%%%%%%
\begin{align}  
    \text{Energy density:} \quad & \rho = T_{ab} u^a u^b = - \tilde{E}, \\  
\text{Null energy density:} \quad & \rho_n = R_{ab} n^a n^b = 2(E + \tilde{E})_{\theta}^{\theta}, \\  
\text{Focusing energy density:} \quad & \rho_f = R_{ab} u^a u^b = E.  
\end{align}  
%%%%%%%%%%%%%%%%%%%%%%%%%%%%%%%%%%%%%%%%%%%%%%%%%%%%%%%%%%%%%%

\noindent In the case of a vacuum, the effective equation is:
%%%%%%%%%%%%%%%%%%%%%%%%%%%%%%%%%%%%%%%%%%%%%%%%%%%%%%%%%%%%%
\begin{equation}  
\rho = \rho_n = 0  
\end{equation}
%%%%%%%%%%%%%%%%%%%%%%%%%%%%%%%%%%%%%%%%%%%%%%%%%%%%%%%%%%%%%
implying
%%%%%%%%%%%%%%%%%%%%%%%%%%%%%%%%%%%%%%%%%%%%%%%%%%%%%%%%%%%%
\begin{equation}  
\tilde{E} = 0, \quad (E + \tilde{E})_{\theta}^{\theta} = 0. 
\end{equation}
%%%%%%%%%%%%%%%%%%%%%%%%%%%%%%%%%%%%%%%%%%%%%%%%%%%%%%%%%%%
It solves to give black hole solutions, Schwarzshild or Kerr. 

\noindent The dual vacuum equation is
%%%%%%%%%%%%%%%%%%%%%%%%%%%%%%%%%%%%%%%%%%%%%%%%%%%%%%%%%%%%%
\begin{equation}  
\rho_f = \rho_n = 0  
\end{equation} 
%%%%%%%%%%%%%%%%%%%%%%%%%%%%%%%%%%%%%%%%%%%%%%%%%%%%%%%%%%%%
leading to
%%%%%%%%%%%%%%%%%%%%%%%%%%%%%%%%%%%%%%%%%%%%%%%%%%%%%%%%%%%%
\begin{equation}  
E = 0, \quad (E + \tilde{E})_{\theta}^{\theta} = 0,  
\end{equation}
%%%%%%%%%%%%%%%%%%%%%%%%%%%%%%%%%%%%%%%%%%%%%%%%%%%%%%%%%%
which solves to the Schwarzschild black hole with a global monopole charge for the static case, while for the stationary case, it is again the Kerr black hole. Unlike the Schwarzschild solution, the Kerr solution is electrograv self-dual. 

\subsection{Vacuum and its dual solutions}

\noindent We consider the axially symmetric spacetime, which will include spherical symmetry when rotation is switched off. We write the metric in the Boyer-Lindquist coordinates, given by  
%%%%%%%%%%%%%%%%%%%%%%%%%%%%%%%%%%%%%%%%%%%%%%%%%%%%%%%%%%%
\begin{equation}\label{Kerr_metric}
ds^2=-\frac{\Delta}{\xi^{2}}\bigg(dt- a \sin^{2} \theta d \phi\bigg)^2+\frac{\xi^{2}}{\Delta} dr^2 + \xi^{2} d\theta^{2}+ \frac{\sin^{2}{\theta}}{\xi^{2}}\bigg(a dt - (r^2+a^2)d \phi\bigg)^{2}~,
\end{equation}
%%%%%%%%%%%%%%%%%%%%%%%%%%%%%%%%%%%%%%%%%%%%%%%%%%%%%%%%%%%%%%%%%%
where $\xi^{2}=r^2+a^2 \cos^2 \theta$. In this gauge, for \ref{Kerr_metric}, $\rho_n = 0$ implying $(E + \tilde{E})_{\theta}^{\theta} = 0$ is automatically satisfied, and hence we need only to solve $\tilde {E} = 0$ for vacuum or ${E} = 0$ for dual vacuum. Let's write $\Delta = r^2f (r) + a^2$ to solve the effective vacuum equation and its dual. For reference, the explicit expressions of the Riemann components are given in \ref{Riemann}. \\

\noindent Now, for the vacuum solution, we have 
%%%%%%%%%%%%%%%%%%%%%%%%%%%%%%%%%%%%%%%%%%%%%%%%%%%%%%%%%%
\begin{equation}
\rho=\tilde{E}= \tilde{E}^{r}_{r} + \tilde{E}^{\theta}_{\theta} +\tilde{E}^{\phi}_{\phi}= -\frac{r^2 \left(r f'(r)+f(r)-1\right)}{\xi^4} =0,
\end{equation}
%%%%%%%%%%%%%%%%%%%%%%%%%%%%%%%%%%%%%%%%%%%%%%%%%%%%%%%%%%%%
\noindent which simply means 
%%%%%%%%%%%%%%%%%%%%%%%%%%%%%%%%%%%%%%%%%%%%%%%%%%%%%%%%%%%%
\begin{equation}
(r(f(r)-1)' = 0;\;\; \textit{i.e.}, \;\;
f(r)= 1- \frac{2M}{r}. 
\end{equation}
%%%%%%%%%%%%%%%%%%%%%%%%%%%%%%%%%%%%%%%%%%%%%%%%%%%%%%%%%%%
\noindent So we obtain the Kerr vacuum solution with 
%%%%%%%%%%%%%%%%%%%%%%%%%%%%%%%%%%%%%%%%%%%%%%%%%%%%%%%%%%%%

\begin{equation}
\Delta= r^2- 2 M r+ a^2.
\end{equation}
%%%%%%%%%%%%%%%%%%%%%%%%%%%%%%%%%%%%%%%%%%%%%%%%%%%%%%%%%%%%
This is all so straightforward and simple.

\noindent Similarly, solving
%%%%%%%%%%%%%%%%%%%%%%%%%%%%%%%%%%%%%%%%%%%%%%%%%%%%%%%%%%%
\begin{eqnarray} \label{global_monopole}
\rho_{f}=E&=& E^{r}_{r}+E^{\theta}_{\theta}+E^{\phi}_{\phi}=\frac{1}{4 \xi ^4} \bigg(2 a^2 r^2 f''(r) \cos^{2}\theta
+4 r f'(r) \left(2 a^2 \cos^2 \theta +r^2\right)\nonumber\\&+&4 a^2 f(r) \cos ^2\theta-4 a^2 \cos^{2}\theta+2 r^4 f''(r)\bigg) = 0~,
\end{eqnarray}
%%%%%%%%%%%%%%%%%%%%%%%%%%%%%%%%%%%%%%%%%%%%%%%%%%%%%%%%%%%%
will give the dual solution, which again turns out to be the Kerr itself. One can write the above equation as 
\begin{eqnarray} \label{global_monopole_1}
\rho_{f}=f_1(r) a^{2} \cos\theta +f_2(r)=0~,
\end{eqnarray}
%%%%%%%%%%%%%%%%%%%%%%%%%%%%%%%%%%%%%%%%%%%%%%%%%%%%%%%%%%%%%\\
where,
%%%%%%%%%%%%%%%%%%%%%%%%%%%%%%%%%%%%%%%%%%%%%%%%%%%%
\begin{equation}
f_{1}(r)=2r^{2}f''(r)+8rf'(r)+4f(r)-4\;\; \textit{and}\; \; f_{2}(r)=4 r^{3}f'(r)+2 r^4f''(r)~.
\end{equation}
%%%%%%%%%%%%%%%%%%%%%%%%%%%%%%%%%%%%%%%%%%%%%%%%%%%%
We observe that there are two parts in the above equation, one free of rotation parameter, $a$, which we denote by $f_2(r)$, and the other by $f_1(r)$. Note that for the static case when $a=0$, we will have $f_2(r)=0$, which reduces to the Laplace equation giving the general solution, $-M/r + const$. This will generate the global monopole. That is how the dual of vacuum in the static spherically symmetric spacetime is the Schwarzschild black hole with a global monopole. 

When $a\neq0$, $f_1(r)=0$ has also to be solved simultaneously. By substituting $f_2(r)=0$ in $f_1(r) = 0$, we arrive at  $(rf(r))' = 0$. That washes out the constant that generates a global monopole. Thus, the dual of the Kerr solution is Kerr itself. It is therefore electrograv self-dual specetime.  

Note that a global monopole is an asymptotic spacetime giving the stresses, $T^{t}_{t}=T^{r}_{r}=\eta^2/r^2$, and the rest being zero. If we solve \ref{global_monopole_1} in the asymptotic limit, then the $f_{1}$ part drops out, and the solution is the same as in the static case with a global monopole. Thus, the Kerr solution is, though, self-dual in general, yet the solution of \ref{global_monopole_1} in the asymptotic limit admits a global monopole. Asymptotically, \ref{global_monopole_1} implies 
%%%%%%%%%%%%%%%%%%%%%%%%%%%%%%%%%%%%%%%%%%%%%%%%%%%%%%%%%%
\begin{equation}
f''(r)+\frac{2 f'(r)}{r} =\nabla^{2}f(r) = 0
\end{equation}
%%%%%%%%%%%%%%%%%%%%%%%%%%%%%%%%%%%%%%%%%%%%%%%%%%%%%%%%%%%%%%
integrating to give 
%%%%%%%%%%%%%%%%%%%%%%%%%%%%%%%%%%%%%%%%%%%%%%%%%%%%%%%%%%%
\begin{equation}
f(r)= 1-\frac{2M}{r}-8 \pi \eta^2,
\end{equation}
%%%%%%%%%%%%%%%%%%%%%%%%%%%%%%%%%%%%%%%%%%%%%%%%%%%%%%%%%%%
\noindent and 
%%%%%%%%%%%%%%%%%%%%%%%%%%%%%%%%%%%%%%%%%%%%%%%%%%%%%%%%%%%
\begin{equation}
\Delta= r^2- 2 M r+ a^2- 8 \pi \eta^2 r^2.
\end{equation}
%%%%%%%%%%%%%%%%%%%%%%%%%%%%%%%%%%%%%%%%%%%%%%%%%%%%%%%%%%%%
\noindent This provides a solution for a Kerr black hole incorporating a global monopole charge asymptotically.

\noindent Not as an electrograv dual, but we shall put a global monopole on a rotating black hole by employing the Newman-Janis transformation to the Schwarzschild black hole with a global monopole. That is what we will do next.

\subsubsection{Newman-Janis Transformation} 

The Newman-Janis algorithm transforms a static black hole into a stationary rotating one; i.e., Schwarzschild to Kerr. We shall apply this algorithm to take Schwarzschild with a global monopole to obtain its rotating analogue, Kerr with a global monopole. 

Here, we apply the Newman-Janis formalism to obtain the metric for the Kerr black hole with a global monopole charge. We start by considering the geometry of a Schwarzschild black hole with a global monopole charge, given by the metric, \cite{Dadhich:1997mh},
%%%%%%%%%%%%%%%%%%%%%%%%%%%%%%%%%%%%%%%%%%%%%%%%%%%%%%%%%%%%%%
\begin{equation}\label{Schwarzschild_global}
ds^{2}=-f(r) dt^{2}+\frac{d r^2}{f(r)}+ r^2(d \theta^{2}+\sin^2\theta)~,
\end{equation}
%%%%%%%%%%%%%%%%%%%%%%%%%%%%%%%%%%%%%%%%%%%%%%%%%%%%%%%%%%%%%%
with 

%%%%%%%%%%%%%%%%%%%%%%%%%%%%%%%%%%%%%%%%%%%%%%%%%%%%%%%%%%%%%%%%%%%%%%%%%%%%%%%%%%%%%%%%%%%%
\begin{eqnarray}
    f &=& \bigg(1-\frac{2M}{r}-8 \pi \eta^2\bigg)~.
\end{eqnarray}
%%%%%%%%%%%%%%%%%%%%%%%%%%%%%%%%%%%%%%%%%%%%%%%%%%%%%%%%%%%%%%%%%%%%%%%%%%%%%%%%%%%%%%%%%%%
By defining a null coordinate as    $dt=du+\frac{dr}{f(r)}$, where $u$, then the metric  \ref{Schwarzschild_global} takes the form,
%%%%%%%%%%%%%%%%%%%%%%%%%%%%%%%%%%%%%%%%%%%%%%%%%%%%%%%%%%%%%%%%%%%%%%
\begin{equation}\label{Schwarzschild}
ds^{2}=-f(r) du^{2}-2 du dr + r^2(d \theta^{2}+\sin^2\theta)~.
\end{equation}
%%%%%%%%%%%%%%%%%%%%%%%%%%%%%%%%%%%%%%%%%%%%%%%%%%%%%%%%%%%%%%%%%%%%%%%
The metric can now be expressed in terms of null tetrads \{l, n, m, $\bar{m}$\}. The vectors $ l$ and $ n$ are real, while $m$ and $\bar{m}$ are complex conjugates where 
%%%%%%%%%%%%%%%%%%%%%%%%%%%%%%%%%%%%%%%%%%%%%%%%%%%%%%%%%%%%%%%%%%%%%
\begin{eqnarray}
l^{a}&=&\delta^{a}_{r}~,\nonumber\\
n^{a}&=&\delta^{a}_{u}-\frac{f(r)}{2}\delta^{a}_{r}~,\nonumber\\
m^{a}&=&\frac{1}{\sqrt{2r}}\left( \delta^{a}_{\theta}-\frac{i}{\sin\theta} \delta^{a}_{\phi}\right)~.
\end{eqnarray}
%%%%%%%%%%%%%%%%%%%%%%%%%%%%%%%%%%%%%%%%%%%%%%%%%%%%%%%%%%%%%%%%%%%%%
It can be shown that $l^{a}n_{a}=-m^{a}\bar{m}_a=1$. Now applying Newman–Janis formalism where $u'\to u-ia \cos \theta$ and $r' \to r+i a \cos \theta$, we arrive at the metric for a Kerr black hole with a global monopole charge in Boyer–Lindquist coordinates. The metric is given by,

%%%%%%%%%%%%%%%%%%%%%%%%%%%%%%%%%%%%%%%%%%%%%%%%%%%%%%%%%%%
\begin{equation}
ds^2=-\frac{\Delta}{\xi^{2}}\bigg(dt- a \sin^{2}d \phi\bigg)^2+\frac{\xi^{2}}{\Delta} dr^2 + \xi^{2} d\theta^{2}+ \frac{\sin^{2}{\theta}}{\xi^{2}}\bigg(a dt - (r^2+a^2)d \phi\bigg)^{2}~,
\end{equation}
%%%%%%%%%%%%%%%%%%%%%%%%%%%%%%%%%%%%%%%%%%%%%%%%%%%%%%%%%%%%%%%%%%
where $\Delta= r^2-2 M r-8 \pi \eta^2 r^2+a^2$.  

This is, however, not a solution to the dual vacuum \ref{global_monopole_1} for arbitrary $r$. On the other hand, for large $r$, when $f_1(r)$ in \ref{global_monopole_1} drops out, it becomes a solution in the asymptotic limit. \noindent The stress-energy tensor components for a Kerr black hole incorporating a global monopole charge are given by,
%%%%%%%%%%%%%%%%%%%%%%%%%%%%%%%%%%%%%%%%%%%%%%%%%%%%%%%%%%
\begin{equation}\label{eq-7}
\begin{split}
  &T_{t}^t= T_{r}^r= \frac{\eta^2 r^2}{\xi^4}\\
&T_{\theta}^{\theta}= T_{\phi}^{\phi}= \frac{\eta^2 a^2 \cos^{2}\theta }{\xi^4}~.
\end{split}
\end{equation}
%%%%%%%%%%%%%%%%%%%%%%%%%%%%%%%%%%%%%%%%%%%%%%%%%%%%%%%%%%%%
Asymptotically, the stress-energy tensors take the following form,
%%%%%%%%%%%%%%%%%%%%%%%%%%%%%%%%%%%%%%%%%%%%%%%%%%%%%%%%%%%%%%%%%
\begin{equation}\label{eq-8}
\begin{split}
   &T_{t}^t= T_{r}^r= \frac{\eta^2}{r^2}\\
&T_{\theta}^{\theta}= T_{\phi}^{\phi}= 0~.
\end{split}
\end{equation}
%%%%%%%%%%%%%%%%%%%%%%%%%%%%%%%%%%%%%%%%%%%%%%%%%%%%%%%%%%%%%%%
\noindent Thus, in the large $r$ limit, we obtain the energy-momentum tensor, which matches that of a Schwarzschild black hole with a global monopole charge \cite{Dadhich:1999eh}. 

%%%%%%%%%%%%%%%%%%%%%%%%%%%%%%%%%%%%%%%%%%%%%%%%%%%%%%%%%%%%%%%%%
%%%%%%%%%%%%%%%%%%%%%%%%%%%%%%%%%%%%%%%%%%%%%%%%%%%%%%%%%%%%%%%%%
\section{Global Monopole}\label{global}

\noindent The simplest choice of a global monopole is a  triplet scalar,
%%%%%%%%%%%%%%%%%%%%%%%%%%%%%%%%%%%%%%%%%%%%%%%%%
\begin{equation}
 \phi ^a = \eta f(r)x^a/\xi~,
 \end{equation}
 %%%%%%%%%%%%%%%%%%%%%%%%%%%%%%%%%%%%%%%%%%%%%%%%%
 where
 $x^a x^a = \xi^{2}$ with the Lagrangian,
 %%%%%%%%%%%%%%%%%%%%%%%%%%%%%%%%%%%%%%%%%%%%
 \begin{equation}\label{lagrangian}
 L=\frac{1} {2}\partial _\mu \phi^a \partial ^\mu \phi^a - \frac{1} {2}\lambda
 (\phi^a \phi^a - \eta ^2)^2~.
 \end{equation}
 %%%%%%%%%%%%%%%%%%%%%%%%%%%%%%%%%%%%%%%%%%%%%%%%%
This model has a global $O(3)$ symmetry which is
spontaneously broken  to $U(1)$.
At large $r$ outside the monopole core, 
 where $f = 1$, 
now, using metric (\ref{Kerr_metric}), we can write the Lagrangian (\ref{lagrangian}) for  the scalar
 field $\phi$ in the following form
 %%%%%%%%%%%%%%%%%%%%%%%%%%%%%%%%%%%%%%%%%%%%%%%%%%%%%%
 \begin{equation}
 L=\frac{\eta ^2 {f^\prime}^2} {2B}+\frac{\eta ^2 f^2} {r^2}-\frac
 {\lambda} {4} \eta ^4 (f^2-1)^2
 \end{equation}
 %%%%%%%%%%%%%%%%%%%%%%%%%%%%%%%%%%%%%%%%%%%%%%%%%%%%%%%%
 where $f^\prime = \partial_r f$ and $B= \frac{\xi^{2}}{\Delta}$. The equation of motion for the field
$\phi$ will be given by
%%%%%%%%%%%%%%%%%%%%%%%%%%%%%%%%%%%%%%%%%%%%%%%%%%%%%%%%%%
\begin{equation}
    B^{-1}\bigg[f^{\prime\prime} + f^\prime \bigg(\frac{A^{\prime}}{2A} -
\frac{B^{\prime}}{2B} + \frac{2}{r}\bigg)\bigg] - \frac{2f}{r^2} - \lambda \eta^2
f(f^2 - 1) = 0~,
\end{equation}
%%%%%%%%%%%%%%%%%%%%%%%%%%%%%%%%%%%%%%%%%%%%%%%%%%%%%%%%%%%%%
where $A= B^{-1}$. This admits an approximate solution $f = 1$ for large $r$ when O($r^{-2})$
is ignorable.\\

\noindent Then $T_{\mu\nu}=2\frac{\partial L}
  {\partial g_{\mu\nu}}-L g_{\mu\nu}$ leads to
 %%%%%%%%%%%%%%%%%%%%%%%%%%%%%%%%%%%%%%%%%%%%%%%%%%%%%%%
 \begin{eqnarray}
T^t_t&=&\frac{\eta ^2 {f^\prime}^2} {2B}+\frac{\eta ^2 f^2} {r^2}+
 \frac{\lambda} {4} \eta ^4 (f^2-1)^2\nonumber\\
 T^r_r&=&-\frac{\eta ^2 {f^\prime}^2} {2B}+\frac{\eta ^2 f^2} {r^2}+
 \frac{\lambda} {4} \eta ^4 (f^2-1)^2\nonumber\\ 
T^{\theta}_{\theta}&=&T^{\phi}_{\phi}=\frac{\eta ^2 {f^\prime}^2}{2B}+
 \frac{\lambda} {4} \eta ^4 (f^2-1)^2 .
 \end{eqnarray}
 %%%%%%%%%%%%%%%%%%%%%%%%%%%%%%%%%%%%%%%%%%%%%%%%%%%%%%%%%
 Now for $r\rightarrow \infty$ and $f=1$ we get
 %%%%%%%%%%%%%%%%%%%%%%%%%%%%%%%%%%%%%%%%%%%%%%%%%%%%%%%%%
 \begin{equation}
T^t_t=T^r_r=\frac{\eta^2} {r^2} .
\end{equation}
%%%%%%%%%%%%%%%%%%%%%%%%%%%%%%%%%%%%%%%%%%%%%%%%%%%%%%%%%%
Asymptotically as $r\to\infty$, the stress energy tensor takes the above form, which matches with that of a Schwarzschild black hole with a global monopole charge spacetime.
%%%%%%%%%%%%%%%%%%%%%%%%%%%%%%%%%%%%%%%%%%%%%%%%%%%%%%%%%%%%%%%%%%%%%%%%%%%%%%%%%%%%%%%%%%%%%%%%%%%%%%%%%%%%%%%%%%%%%%%%%%%%%%%%%%%%%%%%%%%%%%%%%%%%%%%%%%%%%%%%%%%%%%%%%%%%%%%%%%%%%%%%

\section{Discussion} \label{discussion}

The key distinction between Newtonian and Einstein gravity is that in the latter, gravity is self-interactive; i.e., the gravitational field also turns its own source in a subtler way by curving the space without disturbing Newton's inverse square law \cite{Dadhich:2012pda, Dadhich:1997ku}. The Riemann active and passive electric parts, $E_{ab}, \tilde E_{ab}$, are in some way a reflection of these two aspects; $\nabla\Phi(r)$ produced by mass and space curvature by the self-interaction. Electrograv duality characterizes the interchange between them. 

That is, under duality transformation, matter energy $\rho$ goes to focusing energy $\rho_f$. Interestingly, it is the latter that is responsible for the Newtonian gravitational pull. On the other hand, the energy density, $\rho$, characterizes non-empty spacetime in the sense that its absence defines the vacuum spacetime. On the other hand, when $\rho_f=0$, it incorporates Newtonian gravity with $\Phi = -M/r + k$, where $k$ is a constant to be arbitrarily chosen. The difference between the two, $\rho=0$, $\rho_f=0$, is that in the former, $k=0$, implying that the zero of the potential is at infinity and nowhere else, while for the latter, it could be chosen arbitrarily at any $r$ exterior to the gravitating object. It is this $k$ that generates the stresses $T^t_t=T^r_r= k/r^2$ and the rest being zero.  In the asymptotic limit global monopole attains the same stress distribution.

This is how electrograv duality gives rise to a black hole with a global monopole. In the axially symmetric case, the dual vacuum equation $\rho_f=0$ also admits the same vacuum Kerr solution. However, when it is solved in the asymptotic limit, in \ref{global_monopole_1}, the axially symmetric part indicated by the rotation parameter, $a$, drops out, it then reverts to the static solution with a global monopole. 

Thus, the Kerr solution is, in general,  electrograv self-dual; however, asymptotically it could harbor a global monopole like a Schwarzschild black hole. That is, instead of being asymptotically flat, it could have a global monopole asymptotically. By employing the Newman-Janis transformation to the Schwarzschild black hole with global monopole, we obtain the same metric as that obtained by solving the dual vacuum equation asymptotically. 

It is interesting to note that constant potential in the gauge obeying the null energy condition, i.e., $g_{tt}g_{rr}=-1$, produces a topological defect, the solid angle deficit that generates global monopole stresses. There is a novel geometric ansatz to generate a solid angle deficit \cite{Dadhich:1997ue} by writing $4$-dimensional distance proportional to that of $3$-dimensional one; i.e.,  
\begin{equation}
 x_1^2+ ... +x_4^2 = k(x_1^2+ ...+x_3^2).   
\end{equation}
Now, consider a $5$-diemnsional Minkowski spacetime, 
\begin{equation}
 ds^2 = dt^2 - dx_1^2 - ... - dx_4^2.  
\end{equation}
On substituting for $dx_4$ from the previous equation, one obtains a spacetime with solid angle deficit, 
\begin{equation}
 ds^2 = dt^2 - dr^2 - (1-k^2)r^2d\Omega^2.  
\end{equation}
This is also a spacetime of constant potential, generating stresses that correspond to a global monopole asymptotically. That is, constant potential or, equivalently, solid angle deficit is a necessary and sufficient condition for a global monopole. However, in the axially symmetric case, this will not be an electrograv dual spacetime except at the large $r$ asymptotic limit. This is because a global monopole, as constructed in the previous Section, is a radially symmetric static object, and characterized by zero gravitational mass indicated by $\rho_f=0$. 

The electrogravity duality is therefore an interplay between the vacuum and its dual spacetime, which harbors a global monopole asymptotically. The dual spacetime is characterized by the vanishing of the focusing energy density, $\rho + \sigma p_i$, which is equivalently also identified by the constant potential and topological defect, the solid angle deficit. The spacetime it describes has zero gravitational charge \cite{Dadhich:1997ue}. A dual spacetime will always have a solid angle deficit asymptotically describing a global monopole. 

It is interesting to note that the solid angle deficit or global monopole metric was first discovered long back \cite{Dadhich1970} as an example of spacetime having only one Riemann component, $R_{\theta\phi\theta\phi}$ non-zero. It was a tentative attempt to find a minimally curved spacetime, which was remarkably free of Newtonian gravity, signified by the potential being constant. It was much later identified   \cite{PhysRevLett.63.341} with a global monopole. Now we recognise it as an electrograv dual spacetime \cite{Dadhich:1999eh,Dadhich:2000jk}.    

It is straightforward to include electric charge by writing $\rho = Q^2/r^4$ for the charged black hole, giving the usual Reissner–Nordström spacetime, and for its dual, $\rho_f = Q^2/r^4$ will give a charged black hole with a global monopole. Similarly, in the axially symmetric case, we can obtain the charged and rotating Kerr-Newman black hole and its dual by writing $\rho = Q^2/\xi^4$ and $\rho_f = Q^2/\xi^4$, respectively.

%%%%%%%%%%%%%%%%%%%%%%%%%%%%%%%%%%%%%%%%%%%%%%%%%%%%%%%%%%%%%%%%%%%%%%%%%%%%%%%%%%%%%%%%%%%%%%%%%%%%%%%%%%%%%%%%%%%%%%%%%%%%%%%%%%%%%%%%%%%%%%%%%%%%%%%%%%%%%%%%%%%%%%%%%%%%%%%%%%%%%%%%%%%%%%%%%%%%%%%%%%%%%%%%%%%%%%%%%%%%%%%%%

%\section*{Acknowledgements}

\appendix
\labelformat{section}{Appendix #1} 
\labelformat{subsection}{Appendix #1}
%%%%%%%%%%%%%%%%%%%%%%%%%%%%%%%%%%%%%%%%%%%%%%%%%%%%%%%%%%%%%%%%%%%%%%%%%%%%%%%%%%%%%%%%%%%%%%%%%%%
%%%%%%%%%%%%%%%%%%%%%%%%%%%%%%%%%%%%%%%%%%%%%%%%%%%%%%%%%%%%%%%%%%%%%%%%%%%%%%%%%%%%%%%%%%%%%%%%%%%
%%%%%%%%%%%%%%%%%%%%%%%%%%%%%%%%%%%%%%%%%%%%%%%%%%%%%%%%%%%%%%%%%%%%%%%%%%%%%%%%%%%%%%%%%%%%%%%%%%%
\section{The Riemann tensor components} \label{Riemann}

The Riemann tensor components for the metric \eqref{Kerr_metric} can be expressed as,

\begin{align}
% \begin{split}
    &R_{trtr}=\frac{1}{2 \, {\xi}^{6}} \bigg(a^{4} r^{2} \cos^{4}\theta \frac{\partial^2\,f}{\partial r ^ 2} + 4 \, a^{4} r \cos^{4}\theta \frac{\partial\,f}{\partial r} + 2 \, a^{4} \cos^{4}\theta f + 2 \, a^{2} r^{4} \cos^{2}\theta \frac{\partial^2\,f}{\partial r ^ 2} - 2 \, a^{4} \cos^{4}\theta \nonumber\\&+ 4 \, a^{2} r^{3} \cos^{2}\theta \frac{\partial\,f}{\partial r} - 6 \, a^{2} r^{2} \cos^{2}\theta f + r^{6} \frac{\partial^2\,f}{\partial r ^ 2} + 6 \, a^{2} r^{2}\cos^{2}\theta\bigg)  ~,\nonumber\\
     &R_{\theta \phi\theta \phi}=-\frac{{\left(r^{2}-3 a^2 \cos^{2} \theta \right)} r^{2} {\left(f  - 1\right)}}{{\xi}^{6}}  ~,\nonumber\\
      &R_{tr \theta \phi}=-\frac{{\left( 2 \, a^{2} (f -1) \sin^{2}\theta - a^{2} r \frac{\partial\,f}{\partial r} \cos^{2}\theta- r^{3} \frac{\partial\,f}{\partial r} - (2 \, a^{2} - 2 \, r^{2}) (f -1)\right)} a r \cos \theta}{{\xi}^{6}}  ~,\nonumber\\
      &R_{t\theta r \phi}= -R_{t\phi r \theta}=\frac{{\left(a^{2} r \cos^{2}\theta \frac{\partial\,f}{\partial r} + 2 \, a^{2} \cos^{2}\theta f  - 2 \, a^{2} \cos^{2}\theta + r^{3} \frac{\partial\,f}{\partial r} - 2 \, r^{2} (f -1)\right)} a r \cos\theta}{2 \, {\xi}^{6}}~.\nonumber\\
      &R_{t\phi t \phi}= R_{t\theta t \theta}= -R_{r\theta r \theta}=- R_{r\phi r \phi}=\frac{{\left(a^{2} r \cos^{2}\theta \frac{\partial\,f}{\partial r} + 4 \, a^{2} \cos^{2}\theta (f -1)+ r^{3} \frac{\partial\,f}{\partial r}\right)} r^{2}}{2 \, {\xi}^{6}}~.\nonumber
% \end{split}
\end{align}
\bibliography{reference.bib}

\bibliographystyle{./utphys1}

\end{document}